\documentclass[reprint, amsmath,amssymb, aps]{revtex4-1}
\usepackage{graphicx}
\usepackage{float}
\usepackage{dcolumn}
\usepackage{bm}
\usepackage{natbib}
\usepackage{hyperref}
\DeclareMathOperator{\arcsinh}{arcsinh}

\begin{document}

\preprint{APS/123-QED}

\title{Marginally trapped and anti-trapped surfaces for matter evolution in D-dimensions}

\author{Konka Raviteja}
\email{konka.raviteja@gmail.com}
\author{Sashideep Gutti}
\email{sashideep@hyderabad.bits-pilani.ac.in}
\affiliation{Department of Physics, Birla Institute of Technology and Sciences-Pilani \\Hyderabad, India}

\date{\today} 

\begin{abstract}
In this paper, we explore the formation of the marginally trapped and marginally anti-trapped surfaces that arise from the evolution of homogeneous dust in D-dimensions with and without the cosmological constant, this is seen through the analytical expressions for such surfaces. We obtain closed form expressions for the Norm of the normal to the Horizon that decides their causal nature and also look at several interesting features of these surface evolution that are significantly different from the four dimensional counterpart. We obtain the expressions for the Ashtekar-Badrikrihnan's Area-balance law for dynamical horizon (spacelike surface) tailored  for the case of spherically symmetric dust evolution in D-dimensions.
\end{abstract}

\maketitle

\title{Marginally trapped and anti-trapped surfaces for matter evolution in D-dimensions}

\author{Konka Raviteja}
\email{konka.raviteja@gmail.com}
\author{Sashideep Gutti}
\email{sashideep@hyderabad.bits-pilani.ac.in}
\affiliation{Department of Physics, Birla Institute of Technology and Sciences-Pilani \\Hyderabad, India}

\section{Introduction}
The blackhole horizon is a very interesting arena where quantum field theory and general relativity come face to face. The relation between the area of the blackhole horizon and its entropy is one of most important developments of recent times and has been a favourite testing ground for various theories of quantum gravity. A study of the dynamics of the horizon evolution is therefore of great importance since one can track the evolution of quantities like entropy and correlate it with the flux of matter or gravitational waves crossing the horizon. The concept of Trapping Horizons was coined (as against the Event Horizon) to locally track the evolving horizon by Penrose \cite{penrose1965gravitational}. Hayward in his paper \cite{hayward1994general} has refined the concept of trapping horizons based on a 2+2 decomposition framework which introduced various trapped horizons like Future Outer Trapped Horizon (FOTH), Future Inner Trapped Horizon (FITH), Past Outer Trapped Horizon (POTH) and Past Inner Trapped Horizon (PITH). Ashtekar, Badrikrishnan et.al \cite{ashtekar2000generic,ashtekar2000isolated,ashtekar2002dynamical,ashtekar2003dynamical,ashtekar2004isolated} have formulated a closely related notions to Hayward's trapped horizons, which is based on a 3+1 spacetime decomposition framework where they have introduced Isolated Horizons, Dynamical Horizons and Time-like Membranes. The Dynamical Horizon is defined as a space-like hypersurface foliated by 2-Spheres such that the expansion for outgoing ($k^a$) and incoming ($l^a$) null normal are $\Theta_k = 0$ and $\Theta_l < 0$ respectively on every leaf of the folitation. The area of the dynamical horizon is shown to increase monotonically provided the null energy condition is satisfied. If there is no matter flux or gravitational waves crossing the dynamical horizon then it becomes null and is called an Isolated Horizon. Likewise it is shown that one can similarly construct a time-like membrane that arise in cosmological or few cases of gravitational collapse. It is shown in \cite{ashtekar2003dynamical} that in contrast to the Dynamical Horizon case, the area monotonically decreases for a timelike membrane. In the work of Booth et.al \cite{booth2005marginally}, many possible situations where one can find Dynamical Horizons, Time-like Membranes are highlighted and also looked at their causal nature following the prescription given in \cite{dreyer2003introduction} to classify the Horizons as time-like or space-like. Busso in \cite{bousso1999holography} has introduced a construction for Past holographic screen and Future holographic screen which are be defined in terms of Marginally Trapped Surfaces (MTS) or Marginally Anti-Trapped Surfaces (MATS) respectively. Using this construction Busso and Engelhardt in \cite{bousso2015new},\cite{bousso2015proof} have proved a new area law in general relativity where the area of a holographic screen changes monotonically even though the causal nature of the screen (horizon) changes during its evolution.
\\

In this paper, we work with a model \cite{tibrewala2008classical} where the matter content is pressure-less dust in a spherically symmetric arrangement in D-dimensions (the Lemaitre-Tolman-Bondi model generalized to D-dimensions). This model for matter evolution has the advantage that it is general enough to capture many features of Horizon evolution and is simple enough to yield closed form expressions for various scenarios like FOTH, FITH, POTH, PITH. This can therefore yield the D-dimensional versions of various results highlighted in \cite{booth2005marginally},\cite{sherif2019some,helou2017causal,helou2015dynamics}. The area balance law for dynamical horizon given in 3 + 1 dimensions is extended to D dimensions for a spherical topology, which is applied to this model. The analysis of the model in this paper can be used to represent two situations. First, it can represent the evolution of matter of a star if we put a cut-off for the density at some finite radius. Second, it can be interpreted as a cosmological solution (when we set the density to be homogeneous). This analysis is inclusive of the cosmological constant.
\\

For this model, we  obtain the general expression $\Theta_k$, $\Theta_l$ by defining the out-going and in-going null rays in the evolving space-time. Based on the expressions obtained, we can obtain the curve $\Theta_k = 0$ or $\Theta_l = 0$ in the form f(r,t,D,$\Lambda$) = constant in the relevant time and radial coordinate (t, r) respectively. We evaluate the Norm of the vector orthogonal to the these curves and from this deduce the signature of the Horizon. We show that this is equivalent to the prescription given by Booth, et. al \cite{booth2005marginally} where the signature of the curves (Horizons) is evaluated using the ratio of Lie derivatives of $\Theta_k$ and $\Theta_l$. Here we will see that the The causal nature of the D-dimensional horizon evolution is indeed richer and more varied compared to the 3+1 scenario. For e.g it is well known that the for the case of 3 + 1 dimensional Oppenheimer-Snyder matter evolution, the MTS is a time-like membrane. We show in the article that in D-dimensional Oppenheimer-Snyder dust evolution, the MTS is timelike for dimension $D < 5$ and is null for $D = 5$ and is space-like for $D > 5$. The causal nature of the horizon becomes more interesting  when we analyze the MTS, MATS. We observe that in the evolution of MTS and MATS, the Horizon makes a transition from time-like to space-like. We show the expressions highlighting these transitions in the article. We show that even though there is transition from time-like to space-like, the evolution of the Area is strictly monotonic in the time coordinate used in the model (in accordance with the results in \cite{bousso2015new}, \cite{bousso2015proof}). 

\section{Higher dimensional Spherically symmetric dust evolution}
\label{sec:intro}
The general metric for an (D = n+2) dimensional spherically symmetric spacetime is of the form
    \begin{equation}
    ds^2 = - e^{\mu(t,r)}dt^2 + e^{\lambda(t,r)}dr^2 + R^2 (t,r)~d{\Omega}_{n}^{2}
    \end{equation}
where $d{\Omega}_{n}^{2}$ is the metric on unit n dimensional sphere, t is the time coordinate and r is the co-moving radial coordinate. It is easily shown \cite{tibrewala2008classical} that the $g_{00}$ component of the metric can be chosen to be minus one i.e. $g_{00} = -1$. The metric then has the following form.
    \begin{equation}\label{metric}
    ds^2 = -dt^2 + e^{\lambda(t,r)}dr^2 + R^2 (t,r)~d{\Omega}^2
    \end{equation}
For a non-zero cosmological constant ($\Lambda \neq 0$), the Einstein equations are
    \begin{equation}
    G_{\mu\nu} + \Lambda g_{\mu\nu} = \kappa T_{\mu\nu}
    \end{equation}
here $\kappa$ is a constant and is related to Gravitational constant $G_n$, ($\kappa$ = 8$\pi G_n$). The matter we are considering here is a pressure less dust hence the only nonzero component of the stress-energy tensor (in the co-moving and synchronous coordinate system) is $T_{00} = \epsilon (t,r)$, where $\epsilon (t,r)$ is the energy density of the dust. With these conditions we get the Einstein Equations which are shown in \cite{tibrewala2008classical}  and summarized below
    \begin{eqnarray}
    && G_{00} = \frac{e^{-\lambda}}{R^2} \bigg[-\Lambda e^{\lambda} R^{2} + \frac{n(n-1)}{2} (e^{\lambda} (1+\dot{R}^2) - R'^{2}) \nonumber \\ && +  \frac{n}{2} R R' \lambda' +  \frac{n}{2} (-2R R'' + e^{\lambda} R \dot{R} \dot{\lambda}) \bigg] = k \epsilon(t,r)
    \end{eqnarray}
    \begin{equation}
    G_{01} = \frac{n}{2}\frac{(R' \dot{\lambda} - 2 \dot{R}')}{R} = 0
    \end{equation}
    \begin{eqnarray}
    G_{11} && = \frac{1}{R^{2}} \bigg[\frac{n(n-1)}{2} (R'^{2} - e^{\lambda}(1+\dot{R}^2)) \nonumber \\ && + \Lambda e^{\lambda} R^{2} - n e^{\lambda} R \ddot{R}  \bigg] = 0
    \end{eqnarray}
    \begin{eqnarray}
    G_{22} &&= - \frac{1}{4} e^{-\lambda} \bigg[ 2(n-2)(n-1) (e^{\lambda}(1+\dot{R}^2) - R'^{2}) \nonumber \\ && - 2(n-1)(2RR'' - RR'\lambda'- e^{\lambda}(R\dot{R}\dot{\lambda} + 2R\ddot{R})) \nonumber \\ && +  e^{\lambda}R^{2}(-4\Lambda + \dot{\lambda}^2 + 2\ddot{\lambda}) \bigg] = 0 
    \end{eqnarray}
The other non zero relations are given by 
    \begin{equation}
    G_{(j+1~j+1)} = sin^2{\theta_{(j-1)}} G_{(jj)}
    \end{equation}
where j takes values from 2 to n+1. 
\\
The expressions for the evolution of matter can be obtained by simplifying the above set of equations. Solving for the $G_{01}$ we get
    \begin{equation} \label{elambda}
    e^{\lambda} = \frac{R'^{2}}{1 + f(r)}
    \end{equation}
where $f(r)$ is an arbitrary function called the energy function. Integration of the $G_{11}$ equation after using the above relation gives
    \begin{equation}\label{rdotsquare}
    \dot{R}^2 = f(r) + \frac{2\Lambda}{n(n+1)} R^2 + \frac{F(r)}{R^{(n-1)}}
    \end{equation}
where $F(r)$ is called the mass function. Solving for $G_{00}$ we find 
    \begin{equation}\label{density}
    \kappa \epsilon(t,r) = \frac{n F'}{2 R^n R'}
    \end{equation}
This gives us the expression for the mass function  as
    \begin{equation}\label{massfunction}
    F(r) = \frac{2\kappa}{n} \int \epsilon(0,r) r^n dr
    \end{equation}
where $\epsilon(0,r)$ is the initial energy density of the dust and choose that at t = 0, R = r.  We work for the case of marginally bounded shells of dust where we require that $f(r) = 0$. The result (\ref{massfunction}) is obtained by keeping the constant value of f(r) = 0, and this holds true from here on.

\subsection{Solutions of Homogeneous Dust Evolution}
The advantage of the Homogeneous case is that the physical radius R (area radius) is separable into a time dependent part and the co-moving radius r
    \begin{equation}\label{sepofvariables}
    R(r,t) = a(t)~r
    \end{equation}
This model can be used in couple of ways. One is that these solutions can be used to describe cosmological solutions. The other way is that these solutions could represent the generalization of Oppenheimer-Snyder evolution of dust to a general dimension with a cosmological constant. In the latter case the solutions described in this section will work as the interior solutions that need to be matched to an exterior solution (generalization of Schwarzschild solution to higher dimensions including a cosmological constant).

We choose $a(t=0)=1$ thus the co-moving radius $r$ is equal to the physical radius R (area radius) at initial time (t = 0). For a homogeneous dust collapse $\epsilon (t,r)$ is a function only of time $t$ so the initial time density profile of the dust cloud $ \epsilon (0,r) $ is taken to be a positive constant which does not depend on the value of r. So the mass function (\ref{massfunction}) in the homogeneous case is
    \begin{equation} \label{homomassfunction}
    F(r)  = \frac{2g}{n (n+1)}~ r^{n+1}
    \end{equation}
where  $g$ = $\kappa$ $\epsilon (0,r)$ and $g>0$.

Also in the homogeneous case the expression (\ref{rdotsquare}) becomes
    \begin{equation}\label{homordotsquare}
    \dot{R}^2 = \frac{2\Lambda}{n (n+1)} R^2 + \frac{2g~ r^{n+1}}{n (n+1) R^{n-1}}
    \end{equation}
and under the condition of (\ref{sepofvariables}) this reduces to 
    \begin{equation}\label{adotequation}
    \dot{a}(t)^2 = \frac{2\Lambda}{n (n+1)} ~ a(t)^2 + \frac{2g}{n(n+1)~a(t)^{n-1}}
    \end{equation}
In the subsections below, we look at the solutions of (\ref{adotequation}) for various cases of the cosmological constant being zero, negative and positive.
\subsubsection{Case of flat spacetime}
The equation (\ref{adotequation}) for flat spacetime is
    \begin{equation}
    \dot{a}(t)^2 =  \frac{2g}{n(n+1)~ a(t)^{n-1}}
    \end{equation}
For the initial condition that we choose $a(0) = 1$ which means that $R(0,r) = r$, then the solutions for $a(t)$ are
    \begin{equation}\label{flata+}
    \bigg( 1 + \sqrt{\frac{g (n+1)}{2n}}  t  \bigg)^{\frac{2}{1+n}}
    \end{equation}
    and
    \begin{equation}\label{flata-}
    \bigg( 1 - \sqrt{\frac{g (n+1)}{2n}} t  \bigg)^{\frac{2}{1+n}}
    \end{equation}
\subsubsection{Case of Anti de-Sitter spacetime}
The equation (\ref{adotequation}) for Anti de-Sitter spacetime is
    \begin{equation}
    \dot{a}(t)^2 = \frac{- 2\Lambda}{n (n+1)} ~ a(t)^2 + \frac{2g}{n (n+1)~ a(t)^{n-1}}
    \end{equation}
for the initial condition $a(0) = 1$ the solutions for $a(t)$ are 
    \begin{equation}\label{adsa+}
    \bigg(\sqrt{\frac{g}{\Lambda}} \sin{\bigg( \sqrt{\frac{\Lambda (n+1)}{2n}}} t + \arcsin{\sqrt{\frac{\Lambda}{g}}\bigg)} \bigg)^{\frac{2}{n+1}}
    \end{equation}
and
    \begin{equation}\label{adsa-}
    \bigg(-\sqrt{\frac{g}{\Lambda}} \sin{\bigg( \sqrt{\frac{\Lambda (n+1)}{2n}}} t - \arcsin{\sqrt{\frac{\Lambda}{g}}\bigg)} \bigg)^{\frac{2}{n+1}}
    \end{equation}
\subsubsection{Case of de-Sitter spacetime}
The equation (\ref{adotequation}) for de-Sitter spacetime is
    \begin{equation}
    \dot{a}(t)^2 = \frac{2\Lambda}{n (n+1)} ~ a(t)^2 + \frac{2g}{n (n+1)~ a(t)^{n-1}}
    \end{equation}
The solutions of $a(t)$ for our initial condition $a(0) = 1$ are 
    \begin{eqnarray}\label{dsa+}
    \bigg(\sqrt{\frac{g}{\Lambda}} \sinh{\bigg( \sqrt{\frac{\Lambda (n+1)}{2n}}} t + \arcsinh{\sqrt{\frac{\Lambda}{g}}\bigg)} \bigg)^{\frac{2}{n+1}}
    \end{eqnarray}
and
    \begin{eqnarray}\label{dsa-}
    \bigg(-\sqrt{\frac{g}{\Lambda}} \sinh{\bigg( \sqrt{\frac{\Lambda (n+1)}{2n}}} t - \arcsinh{\sqrt{\frac{\Lambda}{g}}\bigg)} \bigg)^{\frac{2}{n+1}}
    \end{eqnarray}

\section{Marginally Trapped and Anti-Trapped Surfaces}
Marginally Trapped Surfaces (MTS) are defined as co-dimension 2 sub-manifolds $\Sigma$ whose expansion of null congruence $\Theta_k$ generated by the outgoing radial null vector $k^a$ vanishes everywhere ($\Theta_k$ = 0) on $\Sigma$ and $\Theta_l$ which is expansion of null congruence generated by incoming radial null vector $l_a$ is completely negative on $\Sigma$ ($\Theta_l < 0$). For these definitions and more see the following references (\cite{ashtekar2002dynamical},\cite{ashtekar2003dynamical},\cite{booth2005marginally},\cite{sherif2019some},\cite{hayward2004energy,boothBHboundary,bengtsson2011region,senovilla1998singularity,senovilla2003existence,senovilla2003novel}).

Similarly we define Marginally Anti-Trapped Surfaces (MATS) as co-dimension 2 sub-manifolds $\Xi$ whose expansion of incoming radial null congruence $\Theta_l$ vanishes everywhere ($\Theta_l$ = 0) on $\Xi$ and the expansion of outgoing radial null congruence $\Theta_k$ is completely positive on $\Xi$ ($\Theta_k > 0$). These MTS and MATS are also referred to  as future and past holographic screens respectively, mostly in the context of holographic theories  (\cite{bousso2015new},\cite{bousso2015proof},\cite{grado2018marginally}).
\\

From the metric (\ref{metric}) we have the future outgoing radial null vector as 
    \begin{equation}
    k^a = (1,e^{-(\frac{\lambda}{2})}, 0, 0, 0,..,0)
    \end{equation}
and the future incoming radial null vector as
    \begin{equation}
    l^a = (1,- e^{-(\frac{\lambda}{2})}, 0, 0, 0,..,0)
    \end{equation}
these two future directed radial null vectors are normalized as
    \begin{equation*}
    k^c l_c = k^c l^d g_{cd} = -2
    \end{equation*}    
and $h_{ab}$ is the induced metric on the marginally trapped or marginally anti-trapped surface which is
    \begin{equation*}
    h_{ab} = g_{ab} + \frac{(k_a l_b + l_a k_b)}{(-k^c l^d g_{cd})} = g_{ab} + \frac{1}{2}(k_a l_b + l_a k_b)
    \end{equation*}
so the expansion for outgoing bundle of null rays is 
    \begin{equation*}
    \Theta_{k} =  h^{ab} \nabla_{a} k_{b} = \frac{n}{R} ( \dot{R} + e^{-(\frac{\lambda}{2})} R' )
    \end{equation*}
using the equation (\ref{elambda}) we have
    \begin{equation}\label{thetak}
    \Theta_{k} = \frac{n}{R} ( \dot{R} + 1 )
    \end{equation}
similarly the expansion for ingoing bundle of null rays is
    \begin{equation*}
    \Theta_{l} =  h^{ab} \nabla_{a} l_{b}
    = \frac{n}{R} ( \dot{R} - e^{-(\frac{\lambda}{2})} R')
    \end{equation*}
again using the equation (\ref{elambda}) we have
    \begin{equation}\label{thetal}
    \Theta_{l} = \frac{n}{R} ( \dot{R} - 1 )
    \end{equation}   

\subsection{Causal Nature of the Marginally Trapped and Anti-Trapped Tubes: General D dimensional LTB}
 The Marginally Trapped Tubes (MTT) or Marginally Anti-Trapped Tubes (MATT) are co-dimension 1 sub-manifolds which are foliated by the marginally trapped or marginally anti-trapped surfaces respectively. We look at the causal nature of these tubes formed in the evolution of the dust, these tubes can be timelike (Timelike tubes), spacelike (dynamical horizons) or null (isolated horizons) depending on various situations which emerge (\cite{booth2005marginally},\cite{sherif2019some},\cite{helou2017causal}). This is done using two methods, in the first one we calculate the norm of the normal to the tubes in (r,t) plane and use it to classify the causal nature of these tubes (applicable for the case of Spherical Symmetry) while the second one is a standard method used (\cite{dreyer2003introduction}) where the ratios of lie derivatives of the expansions are taken which determines the causal nature of the tangent vector to the tube. We use these result to find the causal nature of the tubes that are formed in the dust evolution for the general case
\\~\\
METHOD I:
For this model, we have explicit expression for the tubes in the $(r,t)$ plane. If we consider the expression for $\Theta_{k} = c_1$ where $c_1$ a constant and using (\ref{rdotsquare}) and (\ref{thetak}) expressions we get the curve in the $(r,t)$ plane give by
    \begin{equation}\label{thetaplus}
    \Theta_k =  \frac{n}{R}\left(-\sqrt{\Delta}+1\right) = c_1
    \end{equation}
where
    \begin{equation} \label{Delta}
    \Delta=\frac{2\Lambda}{n(n+1)} R^2 + \frac{F(r)}{R^{(n-1)}}
    \end{equation}
The norm of the normal to the curve in the $(r,t)$ plane determines the causal nature, so the components of the normal to this curve are
    \begin{equation*} \label{nt}
    n_t=-\frac{n\dot{R}}{R^2}\left(1-\sqrt{\Delta}\right)-\frac{n\dot{\Delta}}{2R\sqrt{\Delta}}
    \end{equation*}
and 
    \begin{equation*} \label{nr}
    n_r=-\frac{nR'}{R^2}\left(1-\sqrt{\Delta}\right)-\frac{n\Delta'}{2R\sqrt{\Delta}}
    \end{equation*}
note that for $c_1 = 0$ this curve indicates the marginally trapped tube. This means for evaluating the norm of this normal we impose the conditions $\dot{R}=-1$, $\Delta=1$. The norm works out to be,
    \begin{equation}
    \beta_k = -(n_t)^2+\frac{(n_r)^2}{R'^2}=\frac{n^2}{4R^2}\left(-\dot{\Delta}^2+ \frac{\Delta'^2}{R'^2}\right)
    \end{equation}
and also using (\ref{density}) the norm simplifies as, 
    \begin{equation} \label{betak}
    \beta_k = \frac{4\kappa \epsilon R^2}{n^2}\left(\kappa \epsilon+2\Lambda-\frac{n(n-1)}{R^2}\right).
    \end{equation}
So the norm $\beta_k$ is a function of energy density ($\epsilon$) on the tube, the cosmological constant ($\Lambda$), the area radius ($R$) and the number of dimensions ($n = D - 2$).
\\
Now repeating the same exercise for marginally anti-trapped tubes with  $\Theta_l$ as
    \begin{equation}\label{thetaminus}
    \Theta_l =  \frac{n}{R}\left(\sqrt{\Delta}-1\right) = c_2 
    \end{equation}
where $c_2$ is a constant. The norm evaluates to the same expression given by,
    \begin{equation} \label{betal}
    \beta_l = \frac{4\kappa \epsilon R^2}{n^2}\left( \kappa \epsilon + 2 \Lambda - \frac{n(n-1)}{R^2} \right)
    \end{equation}
\\
Method II:
The standard method for determining the causal nature of marginally (anti) trapped tubes was first discussed in \cite{booth2005marginally} and is done by calculation of ratio of Lie derivatives of $\Theta_k$ and $\Theta_l$. The signature of quantities $\alpha_l$ and $\alpha_k$ discussed below determines the causal nature of the tangent vectors to the MATT and MTT respectively. The lie derivatives for $\Theta_k$ are
    \begin{eqnarray*}
    && \pounds_{k} \Theta_{k} = k^{a} \nabla_{a} \Theta_{k} = \frac{n}{R^2} \bigg[ \frac{R\dot{\Delta}}{2\sqrt{\Delta}} + \frac{R \Delta'}{2R'\sqrt{\Delta}} \nonumber \\ && - (\dot{R}+1)(\sqrt{\Delta}+1) \bigg]
    \end{eqnarray*}
    \begin{eqnarray*}
    && \pounds_{l} \Theta_{k} = l^{a} \nabla_{a} \Theta_{k} = \frac{n}{R^2} \bigg[ \frac{R\dot{\Delta}}{2\sqrt{\Delta}} - \frac{R \Delta'}{2R'\sqrt{\Delta}} \nonumber \\ && - (\dot{R}-1)(\sqrt{\Delta}+1) \bigg]
    \end{eqnarray*}
The causal nature of the marginally trapped tube is determined by the ratio
    \begin{equation}
    \alpha_{k} = \frac{\pounds_{k} \Theta_{k}}{\pounds_{l} \Theta_{k}}
    \end{equation}
which has to be evaluated at $\Theta_{k} = 0$ which implies $\dot{R} = -1$ and $\Delta = 1$, we get
    \begin{equation}\label{alphak}
    \alpha_{k} = \frac{-\kappa \epsilon}{\left(\kappa\epsilon +2\Lambda - \frac{n(n-1)}{R^2}\right)}
    \end{equation}
Similarly the causal nature of the marginally anti-trapped tubes is determined by the the ratio     \begin{equation}
    \alpha_l = \frac{\pounds_{l} \Theta_{l}}{\pounds_{k} \Theta_{l}}
    \end{equation}
which has to be evaluated at $\Theta_{l} = 0$ which implies $\dot{R} = 1$ and $\Delta = 1$, we get the $\alpha_l$ ratio to be 
    \begin{equation}\label{alphal}
    \alpha_l = \frac{-\kappa \epsilon}{\left(\kappa\epsilon +2\Lambda - \frac{n(n-1)}{R^2}\right)}
    \end{equation}
From the equations (\ref{betak}),(\ref{alphak}) it is clear that $\beta_k>0$ is equivalent to $\alpha_k<0$ and imply the MTT is time-like also  $\beta_k<0$ is equivalent to $\alpha_k>0$ and imply the MTT is space-like. Similarly from (\ref{betal}), (\ref{alphal}) when $\beta_l>0$ ($\alpha_l<0$) imply MATT is time-like and $\beta_l<0$ ($\alpha_l<0$) means MATT is space-like. The expressions for $\alpha_k$ and $\alpha_l$ match with the results obtained (equation (23)) in \cite{helou2017causal} and (equation(2.3)) in \cite{booth2005marginally} where the latter expression involves energy density, pressure and the area. We note that the formula depending on the 'area' is valid in four dimensions only and in other dimensions, the expression continues to depend on $R^2$ which does not have the interpretation of 'area' for the MTT or MATT.
\\

We summarize below the Lie derivatives for MTT ($\Theta_k = 0$)
    \begin{equation}\label{lkthetak}
    \pounds_{k} \theta_{k}  = -\kappa \epsilon
    \end{equation}
    \begin{equation}\label{llthetak}
    \pounds_{l} \theta_{k}  = \left( \kappa \epsilon  +2\Lambda -\frac{n(n-1)}{R^2} \right)
    \end{equation}
and MATT ($\Theta_l = 0$)
    \begin{equation}\label{llthetal}
    \pounds_{l} \theta_{l} = -\kappa \epsilon 
    \end{equation}
    \begin{equation}\label{lkthetal}
    \pounds_{k} \theta_{l} =  \left( \kappa \epsilon  +2\Lambda -\frac{n(n-1)}{R^2} \right)
    \end{equation}
The lie derivatives in equations (\ref{llthetak}) and (\ref{lkthetal}) are useful in further characterization of the  MTT and MATT into outer and inner horizons as defined by Hayward \cite{hayward1994general}.  
\subsubsection*{Causal nature of marginally (anti) trapped tubes in homogeneous dust evolution}
Here we adapt the expressions derived for the D-dimensional LTB model for the case of homogeneous dust evolution. The mass function for the Homogeneous case (\ref{homomassfunction}) is of the form 
    \begin{equation} \label{massfunction1}
    F(r)=c_0r^{n+1}
    \end{equation}
where $c_0=2\kappa\epsilon(0,r) /(n(n+1))$. We further can show using the same relation that the energy density,
    \begin{equation} \label{kapppaepsilon}
    \kappa \epsilon(t,r)=\frac{n(n+1)c_0}{2 a^{n+1}}
    \end{equation}
plugging  $\dot{R}^2=1$ in the equation (\ref{rdotsquare}) yields the condition for the Marginally (Anti) Trapped Tubes in the current context to be,
    \begin{equation} \label{homoghorizon}
    r^2\left[\frac{c_0}{a^{n-1}}+\frac{2\Lambda a^2}{n(n+1)}\right]=1
    \end{equation}
The above equation relates the co-moving radius with the scale factor for the horizon.
Now substituting the above relation into the expressions for $\beta_k$ (\ref{betak}) and $\beta_l$ (\ref{betal}) and simplifying yields the relation,
    \begin{equation} \label{normhomogeneous}
    \beta_k = \beta_l = \frac{2\kappa \epsilon}{n}\left(3-n+\frac{2\Lambda R^2}{n}\right)
    \end{equation}
this is the formula for the norm of the normal to the MTT and MATT that occur in homogeneous dust evolution in terms of the energy density on the tube ($\kappa \epsilon$), the dimension of the space-time ($D = n+2$), the cosmological constant ($\Lambda$) and the area radius ($R$) of MTT or MATT. Similarly computing the lie derivatives for the homogeneous case we get
    \begin{equation} \label{homlkthetak}
    \pounds_{k} \theta_{k} = -\kappa \epsilon 
    \end{equation}
    \begin{equation}\label{homllthetak}
    \pounds_{l} \theta_{k}  = \frac{n}{2R^2}\left(3-n+\frac{2\Lambda R^2}{n}\right) 
    \end{equation}
    \begin{equation} \label{homllthetal}
    \pounds_{l} \theta_{l} = -\kappa \epsilon 
    \end{equation}
    \begin{equation} \label{homlkthetal}
    \pounds_{k} \theta_{l} =  \frac{n}{2R^2}\left(3-n+\frac{2\Lambda R^2}{n}\right) 
    \end{equation}
the formula for $\alpha_k$ and $\alpha_l$ is given by
    \begin{equation} \label{homalpha}
    \alpha_k = - \frac{- 2\kappa \epsilon R^2}{n}\left(3-n+\frac{2\Lambda R^2}{n}\right)^{-1} = \alpha_l
    \end{equation}

\subsection{Marginally Trapped Surfaces (MTS)}
The condition for marginally trapped surface (MTS) is 
    \begin{equation}
    \Theta_{k} = 0 ~~~ \text{and} ~~~ \Theta_{l} < 0   
    \end{equation}
from the relations (\ref{thetak}) and (\ref{thetal}) the above conditions imply that
    \begin{equation}
    \dot{R} = -1 ~~~ \text{and} ~~~ \dot{R} < 1
    \end{equation}
so $\dot{R} = -1 $ satisfies both these conditions. This means that for $R(r,t) = r~ a(t)$ we can write co-moving radius r for the MTT as function of time
    \begin{equation}\label{mttcomovrad}
    r_h = \frac{-1}{\dot{a}(t)}    
    \end{equation}
and the aeral radius is expressed as
    \begin{equation}
    R_h(t,r) = - \frac{a(t)}{\dot{a}(t)}
    \end{equation}
From the solutions we got for a(t), we look at the behaviour of MTS as they evolve in time t, for the cases where $\Lambda = 0, >0 ~\text{and}~ <0$
\subsubsection{MTS for flat case}
We choose the solution for a(t) for MTS in the case of $\Lambda = 0$ as
    \begin{equation}\label{mttarealrad}
    a(t) = \bigg( 1 - \sqrt{\frac{g (n+1)}{2n}}  t  \bigg)^{\frac{2}{1+n}}
    \end{equation}
This solution of a(t) is chosen such that $r_h$ and $R_h$ can have positive values for MTS. We are interested in obtaining the curve in the $t-r$ plane for which $\theta_k = 0$. Using the condition (\ref{mttcomovrad}) we obtain the expression for $r_h$ as
    \begin{equation}
    \sqrt{\frac{n (n+1)}{2g}} \bigg( 1 - \sqrt{\frac{g (n+1)}{2n}} t  \bigg)^{\frac{n-1}{n+1}}
    \end{equation}
so we plot $r_h$ versus t for various dimensions (D = 3 to 7) below
    \begin{figure}[H]
    \centering 
    \includegraphics[scale=0.95]{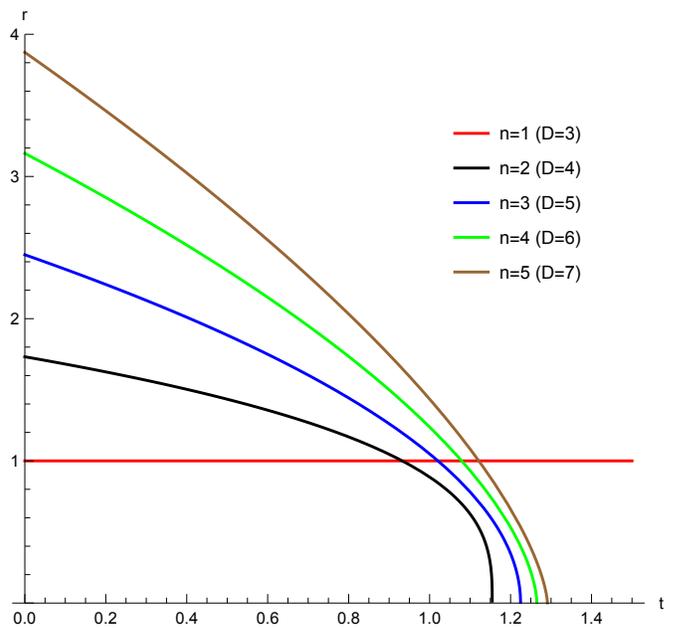}
    \caption{$r_h$ versus t for MTT with ($\Lambda = 0$, g = 1)}
    \label{rvstgeneral}
    \end{figure}
These graphs represent the MTT which gives the evolution of the MTS in time t as each point on these curves is an MTS and we are tracking these MTT curves from time $t=0$ to the time when they reach the shell with label $r=0$ which is a singularity as R also goes to zero here and also the Ricci scalar blows up. We see there is an anomalous curve for $D = 3$ because there are no trapped surfaces in the absence of cosmological constant for the case of $2+1$ dimensions as observed in (\cite{Gutti_2005}). The line for D = 3 represents when the conical defect becomes $2\pi$ in the $2+1$ dimensional scenario and the relation between the conical defect and the mass function $F(r)$ can be seen in (\cite{Gutti_2005},\cite{RossandMann}).
\\

The causal nature of these graphs can be seen from the expression for the norm (\ref{normhomogeneous}), the sign of $\beta_k$ is positive  for $D < 5$ ($n<3$) implying that the MTT is timelike. It becomes null for $ D = 5$ ($n=3$) where the MTT curve coincides with an ingoing null ray, this is an example where the Horizon need not be isolated and can still be null. The MTT is uniformly spacelike for $D > 5$ ($n>3$). We note that for $D = 4$, the MTT is timelike as seen in (\cite{booth2005marginally}, \cite{sherif2019some}). In the above graph and the graphs that follow, we present the evolution for a certain time interval which happens here only due to the matter flux. The MTS unlike the event horizon is defined locally without a need for the complete global description. We look at the evolution for the Areal Radius for these MTS using (\ref{mttarealrad}) which gives us a linear relation between $R_h$ and time (t) given by,
    \begin{equation}
    \sqrt{\frac{n (n+1)}{2g}} - \frac{(n+1)t}{2}
    \end{equation}
The plot for R versus time 
    \begin{figure}[H]
    \centering 
    \includegraphics[scale=0.95]{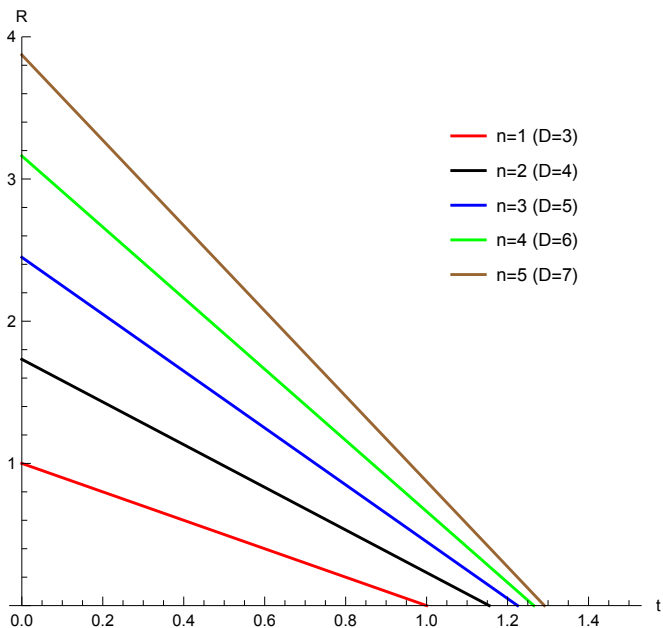}
    \caption{R versus t for MTT with ($\Lambda = 0$, g = 1)}
    \end{figure}
Note that $R$ decreases monotonically with the coordinate time $t$  for all the cases irrespective of the number of dimensions and whether the MTT is space-like or time-like. The areal radius becomes zero at a finite time $t$ indicating singularity formation hence we don't need to extend the solutions beyond $R=0$, this hold true for all the cases to follow.
\\

These MTT can be further characterized as FOTH or FITH as introduced by Hayward \cite{hayward1994general} and outlined in \cite{helou2015dynamics}. Looking at the sign of $\pounds_{l} \Theta_{k} = n\left(3-n\right) / 2R^2$,  we see that for $D>5 (n>3)$ the sign is negative indicating that the horizon is an Outer Horizon (FOTH). For $D<5$, we see that the Horizon is an Inner Horizon (FITH). The non-trivial case is for $D=5$ where the Horizon is null but not isolated. We comment about the Inner and Outer classification for $D=5$ in a note at the end of the section.
\subsubsection{MTS for AdS case}
The solution for the scale factor a(t) in Homogeneous dust evolution for the case with negative cosmological constant is given by
    \begin{equation}
    \bigg(- \sqrt{\frac{g}{\Lambda}} \sin{\bigg( \sqrt{\frac{\Lambda (n+1)}{2n}}} t - \arcsin{\sqrt{\frac{\Lambda}{g}}\bigg)} \bigg)^{\frac{2}{n+1}}
    \end{equation}
One can see that the solutions are oscillatory in nature. The evolution in the graphs given below represent the situation where matter cloud contracts from a given initial configuration and collapses to a point. One can also consider the reversed situation where the matter expands out from a point (this situation is dealt with when analyzing the MATS case). We therefore present a segment of the entire evolution of the cloud for the purpose of tracking the evolution of MTS. The negative cosmological constant provides an extra "attractive force" on the shells and the cloud collapses more efficiently than the previous case where the cosmological constant is kept to zero. The evolution of MTS as a curve in the $(t,r)$ plane where the co-moving radius $r_h$ is given by 
    \begin{equation}
    - \sqrt{\frac{n (n+1)}{2\Lambda}}~\frac{1}{a(t)}~\tan{\bigg( \sqrt{\frac{\Lambda (n+1)}{2n}}} t - \arcsin{\sqrt{\frac{\Lambda}{g}}\bigg)}
    \end{equation}
The plot for $r_h$ versus time t is given by
    \begin{figure}[H]
    \centering 
    \includegraphics[scale=0.95]{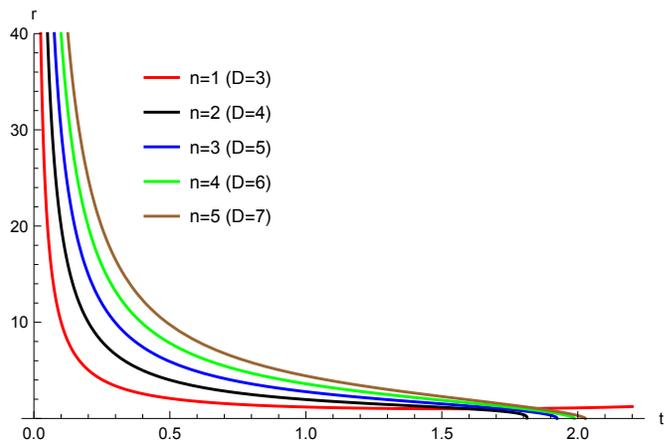}
    \caption{r versus t for MTT with ($\Lambda = 1$, g=1)}
    \end{figure}
For $D \geq 4$, we can see that at time t equal to zero,  the co-moving radius that has just for trapped is the intercept the curve makes on the r axis. For all the co-moving radius $r>r_h$ are already trapped. The evolution therefore proceeds from a higher r to lower r and eventually zero. Once again the case where $D=3$ is anomalous. One can see in the above plot that the MTS never reaches less than a particular value of co-moving coordinate $r$. The reason is that for the case of $2+1$ dimensions not all shells can get trapped (\cite{Gutti_2005}). In the case of Negative cosmological constant, there is a mass gap that needs to be filled before the shells can get trapped. So the shells closer to $r=0$ do not get trapped. In the next plot, we see that all these shells that do not get trapped, do become singular due to their physical radius $R$ becoming zero.
The expression for R is 
    \begin{equation}
    - \sqrt{\frac{n (n+1)}{2\Lambda}}~\tan{\bigg( \sqrt{\frac{\Lambda (n+1)}{2n}}} t - \arcsin{\sqrt{\frac{\Lambda}{g}}\bigg)}
    \end{equation}
and the plot for R versus time  is
    \begin{figure}[H]
    \centering 
    \includegraphics[scale=0.95]{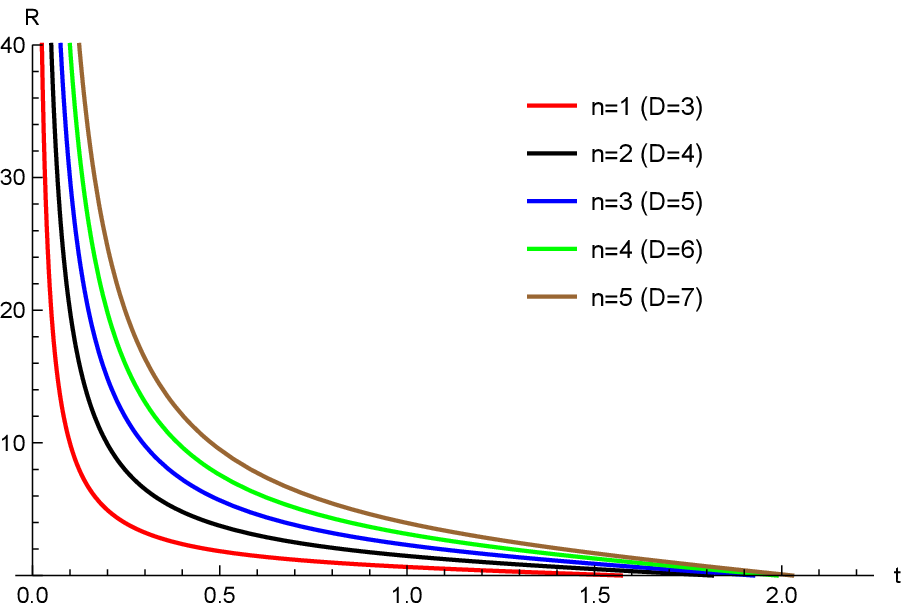}
    \caption{R versus t for MTT with ($\Lambda = 1$, g=1)}
    \label{Rvstmotsadsoplot}
    \end{figure}
As can be seen from the equation (\ref{normhomogeneous}), the MTT is space-like for $D \geq 5$. Time-like behavior is possible only in dimension less than 5. Any MTS that occurs at a areal radius less than $R < \sqrt{n(3-n)/|\Lambda|}$ is time-like and for $R>\sqrt{n(3-n)/|\Lambda|}$ it is space-like. We can see that the above statement is co-variant since the areal radius can be defined in a co-variant manner based on the Killing vectors. We note that the area of the MTS in the negative cosmological constant case decreases monotonically with the coordinate time $t$ for all the cases irrespective of the number of dimensions and whether the MTS is space-like, time-like or mix of time-like and space-like segments.
\\

We now look at the classification of Outer/ Inner based on the expression $\pounds_{l}\Theta_{k} = n\left(3 - n + 2\Lambda R^2/n\right) /2R^2  $. For $\Lambda <0$, it is clear that for $D \geq 5$ the Horizon is uniformly Outer since the above expression is uniformly negative, it is therefore FOTH. For $D<5$ it is an FOTH at large $R$ and is FITH for small $R$. So there is a change from Outer to Inner as the horizon evolves. This counter-intuitive behavior is addressed in the note at the end of the section. 
\subsubsection{MTS for dS case}
The solution for scaling a(t) with positive cosmological constant for MTS is chossen to be (this is choice is made such that r,R are positive)
    \begin{equation}
    \bigg( - \sqrt{\frac{g}{\Lambda}}~\sinh{\bigg( \sqrt{\frac{\Lambda (n+1)}{2n}}} t - \arcsinh{\sqrt{\frac{\Lambda}{g}}\bigg)} \bigg)^{\frac{2}{n+1}}
    \end{equation}
The expression for the MTT curve in (r, t) is plane is give by comoving radius r which is 
    \begin{equation}
    -\sqrt{\frac{n (n+1)}{2\Lambda}}~\frac{1}{a(t)}~\tanh{\bigg( \sqrt{\frac{\Lambda (n+1)}{2n}}} t - \arcsinh{\sqrt{\frac{\Lambda}{g}}\bigg)}
    \end{equation}
The plot for r versus time
    \begin{figure}[H]
    \centering 
    \includegraphics[scale=0.95]{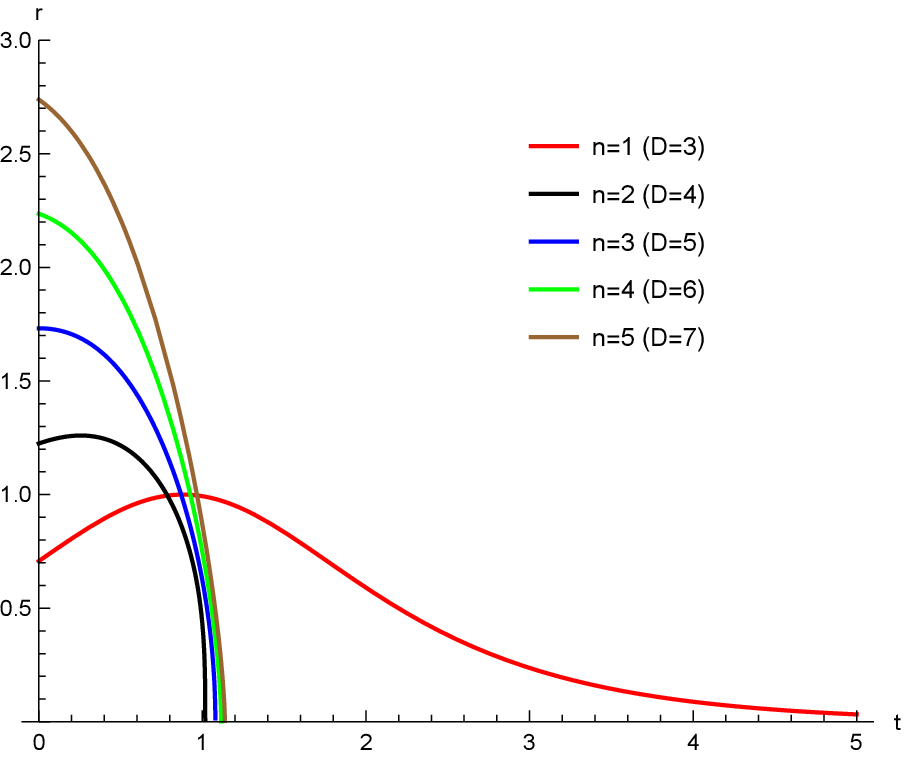}
    \caption{r versus t for MTT with ($\Lambda = 1$, g = 1)}
    \label{rvstmotsds}
    \end{figure}
Just like flat and AdS cases, the curve for the case when $D = 3$ is an anomaly. We look at the expression for the areal radius R of the MTT curve  which is 
    \begin{equation}
    -\sqrt{\frac{n (n+1)}{2\Lambda}}\tanh{\bigg( \sqrt{\frac{\Lambda (n+1)}{2n}}} t - \arcsinh{\sqrt{\frac{\Lambda}{g}}\bigg)}
    \end{equation}
The plot for R versus time 
    \begin{figure}[H]
    \centering 
    \includegraphics[scale=0.95]{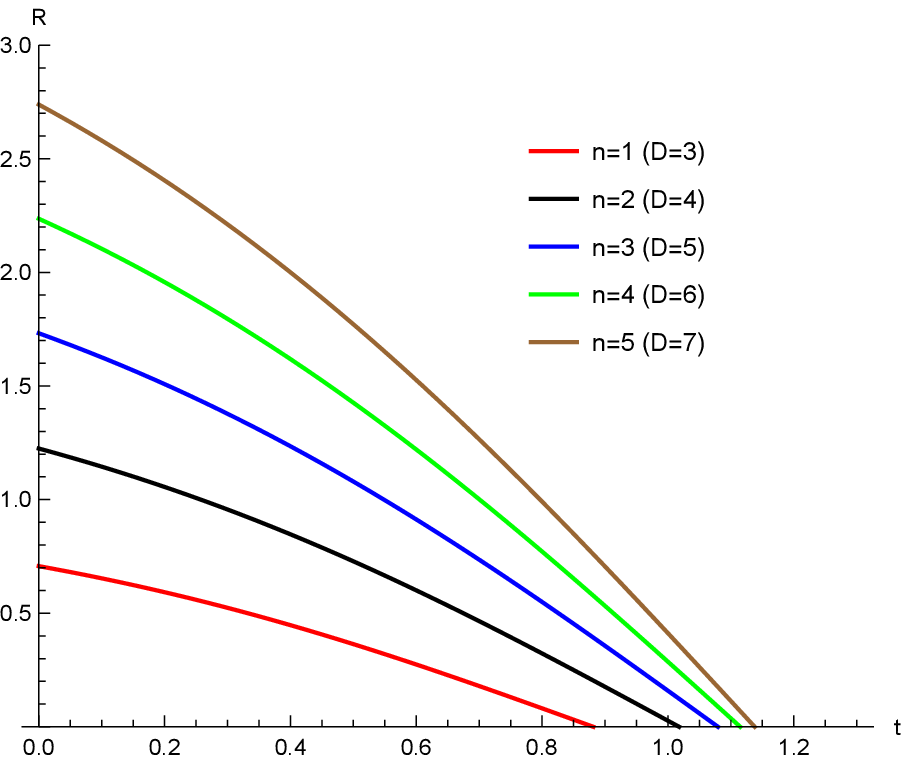}
    \caption{R versus t for MTT with ($\Lambda = 1$, g = 1)}
    \label{Rvstmotsds}
    \end{figure}
We see from the above graphs that just like flat and negative cosmological constant cases, the evolution of MTS for positive cosmological constant case is also monotonic and the areal radius decreases monotonically with time $t$. Using the formula (\ref{normhomogeneous}), we see that for dimensions $d \geq 5$,  the MTS is time-like whenever $R> \sqrt{n(3-n)/(2\Lambda)}$ and is space-like when $R<\sqrt{n(3-n)/(2\Lambda)}$. For dimension $D < 5$, the MTS hyper-surface is time-like since the norm is always positive.
\\

When we analyze the plot ($r,t$) and ($R,t$) together for the case of $D = 3$  dimensions, we see that the portion of the curve in ($r,t$) plot where the slope is positive is the relevant portion. The peak and the downward portion is a result of extending the curve beyond the singularity. This can be seen because when we observe the plot ($R,t$), we can see that the MTS has reached $R=0$ while the curve in the ($r,t$) plot is still climbing. The anomalous behavior of the curve in the ($r,t$) plane is due to the fact that in $D = 3$ dimensions, particles do not attract each other  while the positive cosmological constant has a repulsive effect on the evolving dust. So if a shell of co-moving radius 'r' is such that it's $\dot{R}=-1$ and therefore it is a point on the MTS curve. Due to the repulsive nature of positive cosmological constant, the shell of label $r$ slows down so that it's $\dot{R}>-1$ and a shell with larger co-moving radius will have $\dot{R}=-1$. This explains the peculiar behavior of the $D = 3$ curve. This differs from other dimensions where the evolution of dust is not just dependent on the cosmological constant but also matter distribution that is attractive in nature. 
\\

To characterize the MTS in terms of Outer and Inner, we look at the sign of $\pounds_{l}\Theta_{k} = n\left(3 - n + 2\Lambda R^2/n\right) /2R^2  $. We see that for dimension $D \leq 5$, the sign is uniformly positive implying that the Horizon is an Inner Horizon (FITH). For $D>5$, the Horizon is Inner Horizon (FITH) for large $R$ and is Outer Horizon ( FOTH) for small $R$.

\subsection{Marginally Anti-Trapped Surfaces (MATS)}
The condition for marginally outer trapped surfaces is 
    \begin{equation}
    \theta_{+} > 0 ~~~ \text{and} ~~~ \theta_{-} = 0   
    \end{equation}
these conditions imply
    \begin{equation}
    \dot{R} = 1 ~~~ \text{and} ~~~ \dot{R} > -1
    \end{equation}
when $\dot{R} = 1 $ we can write co-moving radius as function of time as
    \begin{equation}
    r = \frac{1}{\dot{a}(t)}    
    \end{equation}
and the physical radius is expressed as
    \begin{equation}
    R(t,r) = \frac{a(t)}{\dot{a}(t)}
    \end{equation}
we will look at the behaviour of MATS for the cases where $\Lambda = 0, >0 ~\text{and}~ <0$
\subsubsection{MATS for flat case}
The solution for the case of expanding cloud of dust is given by,
    \begin{equation}
    a(t) = \bigg( 1 + \sqrt{\frac{g (n+1)}{2n}} t  \bigg)^{\frac{2}{1+n}}
    \end{equation}
we obtain $r$ as a function of time given by
    \begin{equation}
    \sqrt{\frac{n (n+1)}{2g}} \bigg( 1 + \sqrt{\frac{g (n+1)}{2n}} t  \bigg)^{\frac{n-1}{1+n}}
    \end{equation}
The plot for r versus time for the above relation is given by,
    \begin{figure}[H]
    \centering 
    \includegraphics[scale=0.95]{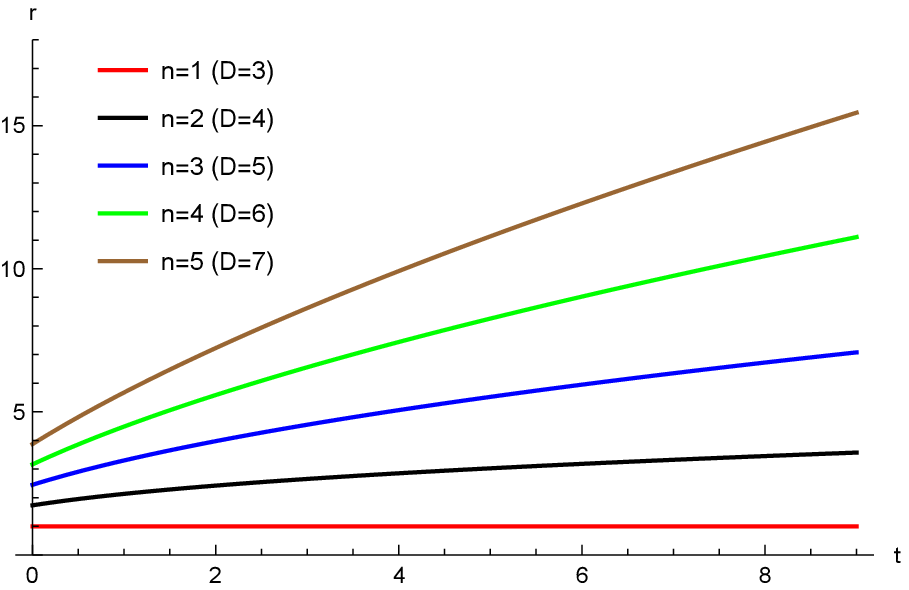}
    \caption{r versus t for MATT with ($\Lambda = 0$, g = 1)}
    \label{rvstmits}
    \end{figure}
Then the expression for the physical radius R for the MATS curve is 
    \begin{equation}
    \sqrt{\frac{n (n+1)}{2g}} + \frac{(n+1)t}{2}
    \end{equation}
The plot for R versus time 
    \begin{figure}[H]
    \centering 
    \includegraphics[scale=0.95]{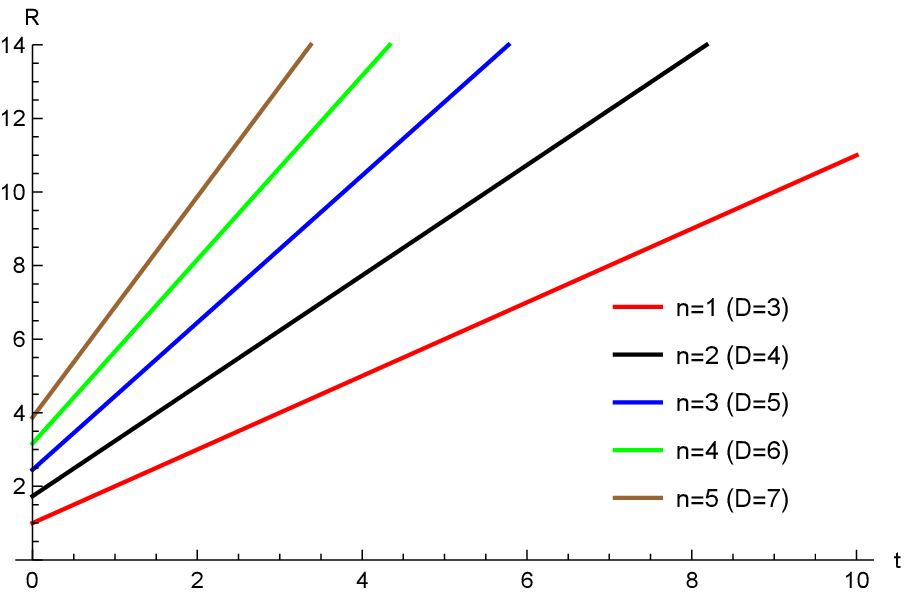}
    \caption{R versus t of MATT with ($\Lambda = 0$, g = 1)}
    \label{Rvstmits}
    \end{figure}
We see a monotonic evolution of $R$ with time. Just as with MTT, the MATT curve is timelike for dimension $D<5$, is null for $D  =5$ and is spacelike for $D>5$.
\\

The MATS could also be further characterized based on the sign of (\ref{homlkthetal}) from which we can see that for $D < 5$ the horizon is Inner (PITH) and for $D > 5$ the horizon is Outer (POTH).
\subsubsection{MATS for AdS case}
The solution for the scale factor a(t) as a function of time for the case with a negative cosmological constant is obtained below (the choice is made such that r,R are positive)
    \begin{equation}
    \bigg( \sqrt{\frac{g}{\Lambda}}~\sin{\bigg( \sqrt{\frac{\Lambda (n+1)}{2n}}} t + \arcsin{\sqrt{\frac{\Lambda}{g }}\bigg)} \bigg)^{\frac{2}{n+1}}
    \end{equation}
The expression for comoving radius r is
    \begin{equation}
    \sqrt{\frac{n (n+1)}{2\Lambda}}~\frac{1}{a(t)}~\tan{\bigg( \sqrt{\frac{\Lambda (n+1)}{2n}}} t + \arcsin{\sqrt{\frac{\Lambda}{g}}\bigg)}
    \end{equation}
The plot for r versus time
    \begin{figure}[H]
    \centering 
    \includegraphics[scale=0.95]{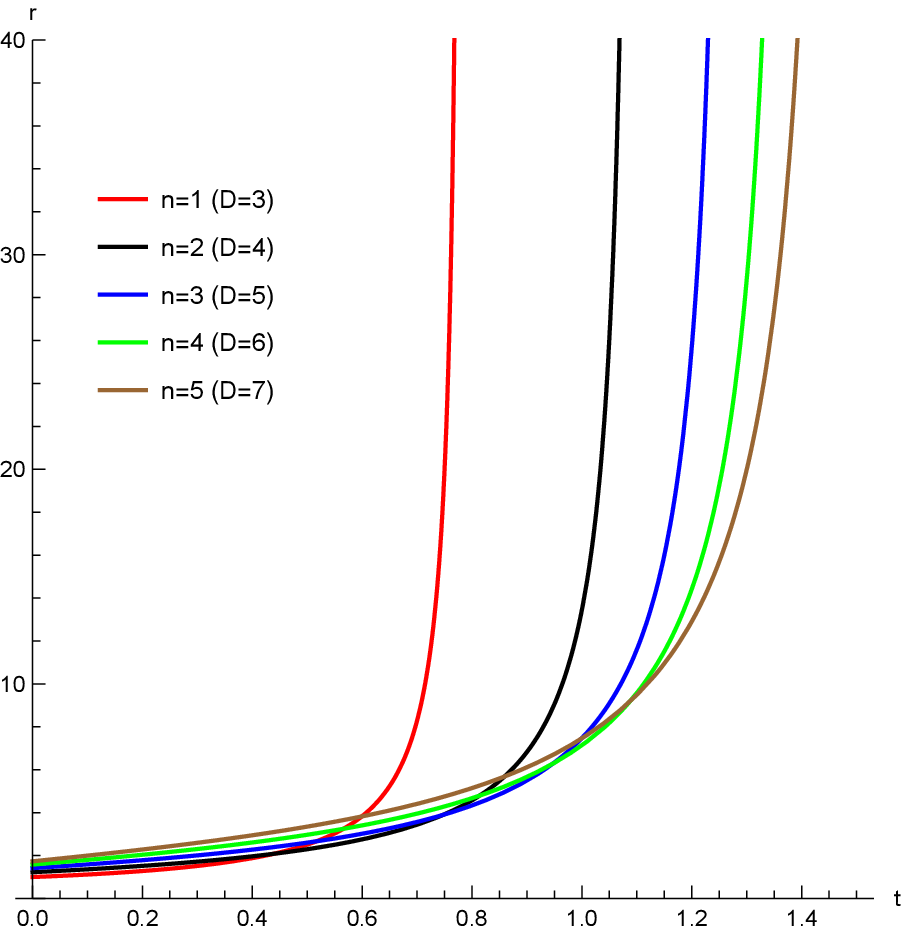}
    \caption{r versus t of MATT with ($\Lambda = 1$, g = 1)}
    \label{rvstmitsinads}
    \end{figure}
The expression for areal radius R is
    \begin{equation}
    \sqrt{\frac{n (n+1)}{2\Lambda}}~\tan{\bigg( \sqrt{\frac{\Lambda (n+1)}{2n}}} t + \arcsin{\sqrt{\frac{\Lambda}{g}}\bigg)}
    \end{equation}
and the plot for R versus time 
    \begin{figure}[H]
    \centering 
    \includegraphics[scale=0.95]{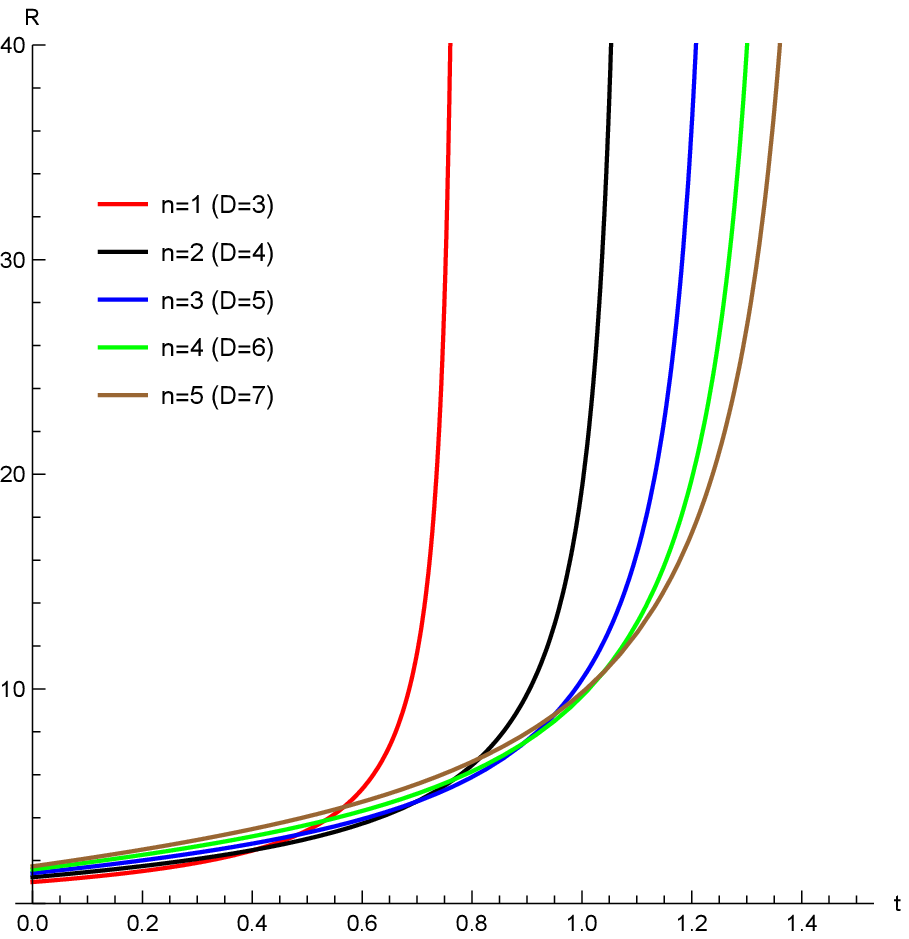}
    \caption{R versus t of MATT with ($\Lambda = 1$, g = 1)}
    \label{Rvstmitsinads}
    \end{figure}
One sees that the solutions $a(t)$ are oscillatory in nature. We consider the expanding part of the solution and track the evolution of MATS. The cloud expands to a maximum and starts contracting back in a finite co-moving time $t$. The steep slope of the MATS curve owes it's explanation to the previous sentence.
\\

Just like MTS,  for dimension $D<5$,  the MATS curve transitions from timelike for small $R$ to spacelike for large $R$. For $D\geq 5$, the curve is uniformly spacelike. When we look at the sign of (\ref{homlkthetal}), we conclude that for $D <5$ the horizon is a PITH for small $R$ and POTH for large $R$ and for dimensions $D \geq 5$, the horizon is POTH.
\subsubsection{MATS for dS case}
The solution for the scaling factor a(t) for MATS in positive cosmological constant is
    \begin{equation}
    \bigg( \sqrt{\frac{g}{\Lambda}}~\sinh{\bigg( \sqrt{\frac{\Lambda (n+1)}{2n}}} t + \arcsinh{\sqrt{\frac{\Lambda}{g}}\bigg)} \bigg)^{\frac{2}{n+1}}
    \end{equation}
The expression for r for MATS  is 
    \begin{equation}
    \sqrt{\frac{n (n+1)}{2\Lambda}}~\frac{1}{a(t)}~\tanh{\bigg( \sqrt{\frac{\Lambda (n+1)}{2n}}} t + \arcsinh{\sqrt{\frac{\Lambda}{g}}\bigg)}
    \end{equation}
and the plot for r versus time is
    \begin{figure}[H]
    \centering 
    \includegraphics[scale=0.95]{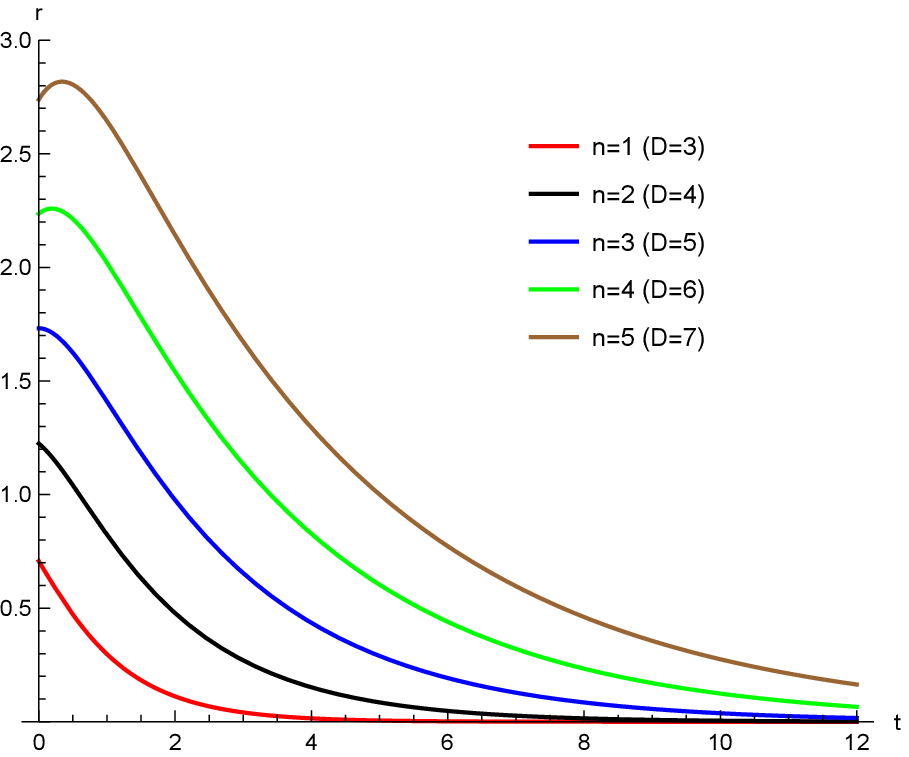}
    \caption{r versus t for MATT with ($\Lambda = 1$, g = 1)}
    \label{rvst}
    \end{figure}
and the expression for areal radius R as a function of time is 
    \begin{equation}
    \sqrt{\frac{n (n+1)}{2\Lambda}} \tanh{\bigg( \sqrt{\frac{\Lambda (n+1)}{2n}}} t + \arcsinh{\sqrt{\frac{\Lambda}{g}}\bigg)}
    \end{equation}
the plot for R versus time for MATS evolution is, 
    \begin{figure}[H]
    \centering 
    \includegraphics[scale=0.95]{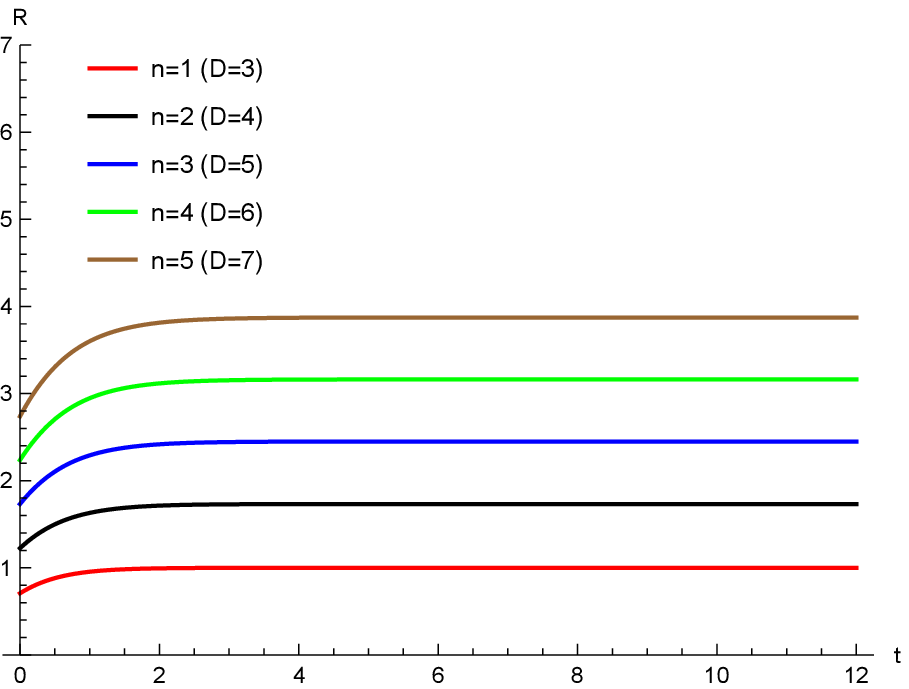}
    \caption{R versus t for MATT with ($\Lambda = 1$, g = 1)}
    \label{Rvstformits}
    \end{figure}
From the expression (\ref{normhomogeneous}) it is clear that if $D \leq 5$ the curve is time-like and is a PITH. As is well known from various work on cosmological horizons in $4$ dimensions regarding horizon evolution due to flux of matter \cite{ashtekar2003dynamical}. For dimension $D>5$ we have MITS curve is space-like (also POTH) for small  $R$ (whenever $R< \sqrt{n(3-n)/(2\Lambda)}$) and is time-like (also PITH) for $R> \sqrt{n(3-n)/(2\Lambda)}$. Just like MTS, there is a transition from space-like to time-like as the MATS evolves. As can be observed from the above plots, $R$ evolves monotonically with time $t$. The plot $R$ vs $t$ is horizontal for large time $t$. This is because most of the matter interior to the MATS surface has crossed the 'cosmological horizon' and therefore the flux of matter is negligible thereof. 

\subsection{Note: Outer and Inner Horizon classification criteria for D = 5 and monotonicity of Horizon evolution}
In this note, we elaborate on two seperate points from the results on the section on MTS and MATS. The first point is the classification of Inner and Outer for the case of $D=5$ with zero cosmological constant. We have seen that the horizon is uniformly null but not isolated. This is seen for both the MTS and MATS case. Now in order to classify the Horizon in terms of being Outer or Inner, we can check with the Lie derivatives $\pounds_{l} \Theta_{k}$ and $\pounds_{k} \Theta_{l}$ (\ref{homlkthetak}, \ref{homllthetak}, \ref{homllthetal}, \ref{homlkthetal}), both are indeed zero. In such situations, we propose another criteria to characterize the Outer or Inner nature of the horizon.
\\

The classification is made possible by defining a space-like vector given by $\epsilon = {\partial}/{\partial R}$ in a coordinate chart $(t,R, \theta, \phi)$ where $t$ is the co-moving time. The vector ${\partial}/{\partial t}$ is everywhere time-like  a good time coordinate  everywhere except at curvature singularity. Now $\epsilon$ points in the direction of increasing areal radius. We can now construct the quantities,  $\pounds_{\epsilon} \Theta_{k}$ for MTS and $\pounds_{\epsilon} \Theta_{l}$ for MATS. The Horizon is Outer if $\pounds_{\epsilon} \Theta_{k}$  is positive and Inner if $\pounds_{\epsilon} \Theta_{k}$ is negative for MATS. Similarly, the horizon is Outer if $\pounds_{\epsilon} \Theta_{l} < 0$ and Inner if $\pounds_{\epsilon} \Theta_{l} > 0$. The use of the vector $\epsilon$ is limited to the case when the MTT or MATT is null. The reason for this is that if the horizon is for space-like, then the expression of the type $\pounds_{\epsilon} \Theta_{k}$ can have variable sign (depending on the time coordinate being used) even though $\pounds_{l} \Theta_{k}$ has an invariant sign.
\\

Evaluating the expression $\pounds_{\epsilon} \Theta_{k}$, we obtain the general expression for $D=5$ with zero cosmological constant,
    \begin{equation}
    \pounds_{\epsilon} \theta_{k}=\frac{3}{R^2}\left[2-\frac{F'}{R'R}\right]
    \end{equation}
Using the results (\ref{massfunction1}), (\ref{homoghorizon}) for $D=5$ we obtain, $\pounds_{\epsilon} \theta_{k}=-6/R^2$, which means we have an Inner Horizon. Similarly, we can classify MATS for the $D=5$ as an Inner horizon. We note that we can reach the same conclusion if we used the space-like vector field to be $\partial/\partial r$ in the co-moving coordinate chart $(t,r,\theta, \phi)$ coordinates system (with the additional assumption that $R'>0$ and therefore $r$ and $R$ are monotonically increasing functions of each other).
\\

The second point that is the observation that in the presence of a cosmological constant, the horizon evolution makes a transition from time-like to space-like as is observed in deSitter case for dimension $D >5$. We note that for larger $R$, the horizon is time-like and small $R$ it is space-like. What seems non-trivial in these cases is that when one classifies the horizon as an Inner or Outer Horizon, one encounters the following situation that as long as the curve is time-like the Horizon is an inner surface whereas in the space-like segment it is an outer surface. When we see the plots for the horizon evolution, we see a monotonic decrease in the areal radius $R$ for both the time-like and space-like segments of the curve. The transition from inner surface to outer surface is counter-intuitive. The analysis of such curves has been done rigorously in \cite{bousso2015new},\cite{bousso2015proof}. The understanding is that for the space-like segment of the curve that decreases monotonically with the coordinate $t$, one can show that for a different choice of coordinates, the time ordering of the events of the space-like segment can be reversed. So in this coordinate system, the space-like part of the horizon evolves from $R=0$ with an increasing Area (since it is OUTER and therefore FOTH) and meets the timelike segment of the curve at the radius $R$ (where the curves in the plots transitions from time-like to space-like). 


\section{Area Laws in Marginally trapped surfaces}
The Area law for a Dynamical horizon (co-dimension 1 spacelike surface) and a Time-like membrane (co-dimesion 1 timelike surface) in 3+1 dimensions is given by Abhay Ashtekar and Badri Krishnan \cite{ashtekar2002dynamical},\cite{ashtekar2003dynamical}. They obtained an area balance law for the dynamical horizon which is 
    \begin{eqnarray}
    \left( \frac{R_2}{2G} - \frac{R_1}{2G} \right) = \int_{\Delta H} \Bar{T}_{ab} \hat{\tau}^a \xi_{(r)}^b d^{3}v && \\ \nonumber + \frac{1}{16 \pi G} \int_{\Delta H} N_r (|\sigma^{2}| + 2|\zeta|^2)d^3v
    \end{eqnarray}
The two terms on the right hand side are the matter energy flux and the gravitational energy flux along the evolution vector $\xi_{(r)}^b$ and similarly the area balance law for the time-like membrane is
    \begin{eqnarray}
    \left( \frac{R_2}{2G} - \frac{R_1}{2G} \right) = - \int_{\Delta H} \Bar{T}_{ab} \hat{r}^a \xi_{(t)}^b d^{3}v && \\ \nonumber - \frac{1}{16 \pi G} \int_{\Delta H} N_t (|\sigma^{2}| - 2|\zeta'|^2)d^3v
    \end{eqnarray}
Using these area laws they have also argued that the area increases for dynamical horizons and decreases for Time-like membrane monotonically.
\\

Following there derivation of Area laws closely we look to extend these laws for Marginally (Anti) Trapped Tubes which are spacelike co-dimension 1 hyper-surfaces to a higher dimensional spacetime (D = n+2) with a topology of $\mathbb{R}^2 \times \mathbb{S}^d$. For a codimension-1 foliation of the spacetime, specifying the evolution vector field $\xi^a$ will also specify the lapse function and shift vectors in the 1+(n+1) decomposition.
    \begin{equation}
    N \tau^a + N^a = \xi^a
    \end{equation}
H is a MTT and is a codimension-1 hypersurface with the Cauchy data and there constraint equations are
    \begin{equation}\label{scalarconstraint}
   C (q,k) :=  R + k^{2} - K^{ab}K_{ab} = 2\kappa \Bar{T}_{ab} \tau^{a} \tau^{b}
    \end{equation}
    \begin{equation}\label{vectorconstraint}
   C^{a}(q,k) :=  D_{b}(K^{ab} - Kq^{ab}) = \kappa \Bar{T}^{bc} \tau_{c} q^{a}_{b}
    \end{equation}
Where $\Bar{T}_{ab} = T_{ab} - (\Lambda g_{ab}/\kappa)$ and $\tau^a$ is the unit normal to H. To get the flux through the a region of MTT ($\Delta H$) bounded by two marginally trapped surfaces at different times we need to evaluate
    \begin{equation}
    \int_{\Delta H} (NC + N_{a}C^{a})d^{n+1}v
    \end{equation}
using the equations (\ref{scalarconstraint}), (\ref{vectorconstraint}) we have
    \begin{eqnarray}
    \int_{\Delta H} (NC + N_{a}C^{a})d^{n+1}v  = 2\kappa \int_{\Delta H} ( N \Bar{T}_{ab} \tau^{a} \tau^{b}  && \\ \nonumber + 2 N_{a} \Bar{T}^{bc} \tau_{c} q^{a}_{b} ) d^{n+1}v
    \end{eqnarray}
For a MTT the choice of the evolution vector field is $\xi^a = N k^a $ and a further 1 + n decomposition of H with $\Sigma$ as a MTS which is a codimension-2 hypersurface with a topology of $\mathbb{S}^n$. With this setup and following the steps as in \cite{ashtekar2002dynamical} we end up with a similar equation as (3.21) in \cite{ashtekar2002dynamical}
    \begin{eqnarray}\label{generalal}
    \int_{\Delta H} N\Tilde{R} d^{n+1}v  = 2\kappa \int_{\Delta H} \Bar{T}_{ab} \xi^{a} \tau^{b} d^{n+1}v && \\ \nonumber + \int_{\Delta H} N (|\sigma|^2 + |\zeta|^2 ) d^{n+1}v
    \end{eqnarray}
The quantities in the gravitational flux energy term are defined below. The shear for the outgoing bundle of light rays is
    \begin{equation}
    \sigma_{ab}^{k} = (h_{a}^{c} h_{b}^{d} - \frac{1}{2}h_{ab}h^{cd}) \nabla_{c} k_{d}
    \end{equation}
where $h_{b}^{a}$ is the projection operator onto the n-sphere and is given by
    \begin{equation}
    h_{b}^{a} = \delta_{b}^{a} - l_b l^a - k^a k_b
    \end{equation}
computing the norm of the shear for outgoing null rays we get
    \begin{equation}\label{sheark}
    |\sigma^{k}|^2 = \sigma_{ab}^{k}~{\sigma^{k}}^{ab}
    = \frac{n(n-2)^2}{4} \frac{e^{-\lambda} ( R' + e^{(\frac{\lambda}{2})} \dot{R} )^2}{ R^2}
    \end{equation}
and using the condition (\ref{thetak}) for MTS we can see that the norm of the shear for outgoing null rays is zero. Similarly the shear for the ingoing bundle of null rays is
    \begin{equation}
    \sigma_{ab}^{l} = (h_{a}^{c} h_{b}^{d} - \frac{1}{2}h_{ab}h^{cd}) \nabla_{c} l_{d}
    \end{equation}
and computing the norm of the shear for ingoing null rays we get
    \begin{equation}\label{shearl}
    |\sigma^{l}|^2 = \sigma_{ab}^{l}~{\sigma^{l}}^{ab}
    = \frac{n(n-2)^2}{4} \frac{e^{-\lambda} (R' - e^{(\frac{\lambda}{2})} \dot{R} )^2}{ R^2}
    \end{equation}
and using the condition (\ref{thetal}) for MATS we see that the shear norm go to zero. The quantity $\zeta$ for MTT is given by the expression
\begin{equation}
    \zeta^a = s^{ab} r^a \nabla_c k_b
\end{equation}
where $s^{ab}$ is the intrinsic metric on MTS. One can easily check that for a spherical symmetry the norm $|\zeta|^2$ is always zero for both the MTS and the MATS. We can see that for spherical dust evolution the gravitational wave energy term always vanishes. Hence the only contribution for the change in marginally trapped surfaces comes from the matter energy flux.
\\

The volume element on the MTT (H) can be written as $d^{n+1}v = N^{-1} dR d^{n}v$ so the expression (\ref{generalal}) reduces to
    \begin{equation}\label{areaequation}
    \int^{R_2}_{R_1} dR \oint_{S^n} \Tilde{R} d^{n}v = 2\kappa \int_{\Delta H} ( N \Bar{T}_{ab} l^{a} \tau^{b}) d^{n+1}v
    \end{equation}
The n-dimensional volume element on $\mathbb{S}^n$ is $d^{n}v = R^{n} sin{\theta_1} sin^2{\theta_2}....sin^{n-1}{\theta_{n-1}} d{\theta} d{\theta_1} d{\theta_2}....d{\theta_{n-1}}$ and Ricci scalar $\Tilde{R}$ for the n-sphere is $\Tilde{R} = {n (n-1)}/{R^2}$, where $R$ is the Areal Radius. So volume integral of Ricci scalar for the n-sphere is
    \begin{equation}
    \oint_{S^n} \Tilde{R} d^{n}v = \frac{2 \pi^{(\frac{n+1}{2})} n (n-1) R^{n-2}}{\Gamma(\frac{n+1}{2})}
    \end{equation}
and the area of the n-sphere with radius R is given by
    \begin{equation}
    A(R) = \frac{2 \pi^{(\frac{n+1}{2})} R^n}{\Gamma(\frac{n+1}{2})}
    \end{equation}
The left hand side of the integral (\ref{areaequation}) becomes
    \begin{equation}
    \int^{R_2}_{R_1} dR \oint_{S^n} \Tilde{R} d^{n}v = \frac{2 \pi^{(\frac{n+1}{2})} n }{\Gamma(\frac{n+1}{2})} (R_{2}^{n-1} - R_{1}^{n-1})
    \end{equation}
For evaluating the matter flux term of the equation (\ref{areaequation}) we use the relation $d^{n+1}v = N^{-1} dR d^{n}v$  again and also (\ref{density}) which simplify the expression as
    \begin{eqnarray}
    2 \kappa \int_{\Delta H} ( N ( T_{ab} - \frac{\Lambda~g_{ab}}{k} ) l^{a} \tau^{b}) d^{n+1}v  && \\ \nonumber = 2 \frac{2 \pi^{(\frac{n+1}{2})}}{\Gamma(\frac{n+1}{2})} \int_{\Delta R}  ( \frac{n F'}{2 R^{n} R'} + \Lambda ) R^{n} dR 
    \end{eqnarray}
now the expression (\ref{areaequation}) reduces to the form
    \begin{align*}
    \int_{\Delta R} n (n-1) R^{n-2} dR - 2 \Lambda \int_{\Delta R} R^{n} dR & = n \int_{\Delta r} F'dr
    \end{align*} 
upon integration have the relation
    \begin{equation}
    (R_{2}^{n-1} - R_{1}^{n-1}) - \frac{2\Lambda}{n (n+1)} (R_{2}^{n+1} - R_{1}^{n+1}) = F(r_2) - F(r_1) 
    \end{equation}
which is
    \begin{equation}
    \Delta R^{n-1} - \frac{2\Lambda}{n (n+1)} \Delta R^{n+1} = \Delta F(r)
    \end{equation}
This is same as the relation (\ref{rdotsquare}) under the marginally (anti) trapped condition $\dot{R}^2 = 1$ which is either $\dot{R} = -1$ $(\Theta_k = 0 )$ and $\dot{R} = 1$ $(\Theta_l = 0 )$. Note that the area balance law has been reduced to an algebraic relation between the misner-sharp mass F(r) and the Area Radius R. Also this extension to D-dimensional area balance law is done for only spacelike MTS or MATS.
\\

\section{Conclusions}
We have generalized the evolution of MTS and MATS in D-dimensions with and without the cosmological constant due to the evolution of pressure-less matter. The model under consideration is simple enough to yield closed form expressions for various aspects of the horizon evolution in these space-times. This advantage makes this model particularly useful in the study of Entropy evolution and Quantum Gravity scenarios. Particularly interesting result among them is the formula for the causal nature of the horizon. The formula highlights the dependence on dimension, local energy density, cosmological constant and the Area radius (D-dimensional generalization of area).
\\

The analysis of MTS  and MATS in D-dimensions yields many interesting results that are not a straightforward extension of the results of $3+1$ dimensions. We observe that the qualitative features of the dynamics of the horizons depends crucially on the number of dimensions D. In the examples that were shown, there were cases where MTS and MATS were uniformly null but not isolated and moreover the area evolves  monotonically. We have also shown that the generalisation of Oppenheimer-Snyder model in D-dimensions yields the Horizon to be time-like for dimension $D<5$ and is space-like for $D>5$. This is interesting since the area monotonically decreases with co-moving time in-spite of the horizon being space-like, time-like in different segments of the same curve.  These results make the analysis in d-dimensions counter-intuitive.
\\

We have founnd expressions for the  Ashtekar, Badrikrishnan's Area Balance Law in D-dimensions for a restricted class of $S^n$ topologies that are relevant for the model under consideration. The expressions obtained in the article are valid for the Dynamical Horizons. One can extend the expressions for the case of Time-like Membranes too (as is shown in the paper Ashtekar et al \cite{ashtekar2003dynamical}). In the cases considered we show that the horizon transitions from time-like to space-like during the course of evolution. The generalization of the $3+1$ of \cite{ashtekar2003dynamical} where such transitions are allowed will be attempted in a later work.

\bibliographystyle{unsrt}
\bibliography{references}

\end{document}